# Design of a 325MHz β=0.12 superconducting single spoke cavity

# for China-ADS∗

LI Han(李菡)[1;2]    DAI Jianping(戴建枰)[2;1)]    SHA Peng(沙鹏)[2]
HUANG Hong(黄泓)[2] WANG Qunyao(王群要)[2]  Zhangjuan(张娟)[1; 2]
YAO Zhongyuan（姚中元）[3] LU Xiangyang（鲁向阳）[3]

1 (Graduate University of the Chinese Academy of Sciences, Beijing 100049, China)
2 (Institute of High Energy Physics, CAS, Beijing 100049, China)
3 （State Key Laboratory of Nuclear Physics and Technology, Peking University, Beijing 100871, China）

**Abstract**：Twelve superconducting single spoke cavities whose Beta is 0.12 (Spoke012) operating at 325MHz, are adopted in Injector I for China-ADS linac. This type of spoke cavity is believed to be one of the key challenges for its very low geometry Beta. So far, the prototype cavity has been designed, fabricated and tested successfully. The design work was finished by using CST-MWS and ANSYS software for the RF and mechanical properties optimization. This paper presents the details of the final design for Spoke012 prototype cavity.

**Key words**: Low Beta spoke cavity, EM design, CST-MWS, ANSYS

PACS: 29.20.Ej

## 1. Introduction

A spoke cavity is a TEM-class superconducting resonator with excellent RF performances in the low and middle Beta region. Besides the inherent advantages of superconductors, the spoke cavity has smaller dimensions and a stronger mechanical structure than the elliptical cavity with the same frequency, and higher R/Q values than the half-wave resonator[1]. For many advantages in the low energy section, it has become a potential candidate for worldwide high intensity accelerators. The project-X under development at Fermilab, a multi-MW proton source, includes three types of SC single spoke cavities operating at 325MHz[2].The

ESS(European Spallation Source) will employ 28 spoke cavities at 352.21MHz[3].The HINS (Fermilab High Intensity Neutrino Source) will use single spoke cavities at 325MHz with β of 0.21[4].

Therefore, the China-ADS (Accelerator Driven Sub-critical System), based on a 1.5GeV CW linac, is also proposed to adopt spoke cavities in low Beta regions. The Injector I of China-ADS linac is shown in Figure 1[5]. The normal RFQ cavity accelerates the beam to about 3.2MeV. Three types of superconducting spoke cavities with operating frequency of 325MHz, Spoke012 at β=0.12, Spoke021 at β=0.21 and Spoke040 at β=0.40, are used to accelerate the beam from 3.2

∗Supported by the "Strategic Priority Research Program" of CAS, Grant No. XDA03020600
1) E-mail: jpdai@ihep.ac.cn

MeV to 178 MeV. Then the beam is accelerated up to ~1.5GeV by two types of superconducting elliptical cavities working at 650MHz.

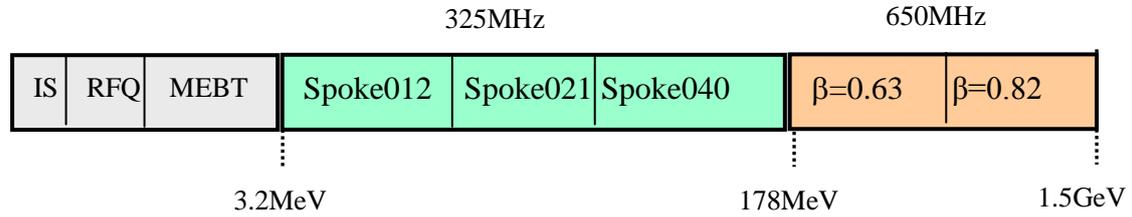

Figure 1: Layout of China-ADS linac with Injector-I

In these three kinds of spoke cavities, Spoke012, which has the smallest geometry Beta, is the first superconducting spoke resonator developed for China-ADS. Its iris length is only 73 mm and the frequency of the small resonance region is very sensitive to elastic end-wall deformation. Consequently, Spoke012 is considered as one of the key challenges for its high value of pressure sensitivity.

Based on the urgent need of ADS project, CW spoke cavity has been developed since 2011. The RF design and mechanical optimization of Spoke012 were finished by IHEP and PKU collaboration group, and the details of design are discussed in this paper.

## 2. RF design

Required by beam dynamics, the frequency of 325MHz, $\beta$ of 0.12 and beam aperture diameter of 35mm are chosen for Spoke012 cavity, and the iris-to-iris distance is defined to be $2/3\beta\lambda$.

In order to get a higher accelerating gradient, the RF parameters of Spoke012 cavity are optimized by maximizing the shut impedance and minimizing the ratio of $E_{peak}/E_{acc}$ and

$B_{peak}/E_{acc}$. The RF design and optimization of Spoke012 has been done by CST_MWS (Microwave studio) software. Many geometrical parameters of cavity could be changed to optimize the Electromagnetic field, as shown in Figure 2.

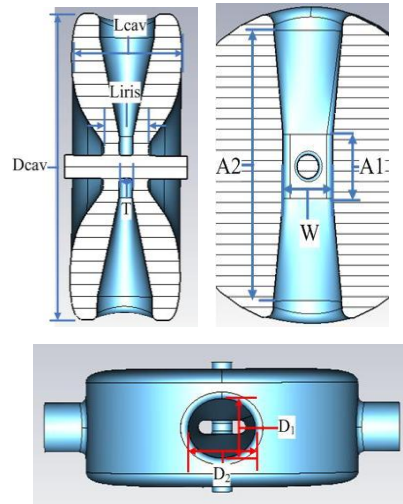

Figure 2: Cross section of spoke012 cavity ($L_{cav}$-cavity length, $L_{iris-iris}$ to iris length, T-spoke thickness, W-spoke width, $D_{cav}$-cavity diameter, D1-minor axis of spoke base, D2- major axis of spoke base.).

The electromagnetic profile is sensitive to the shape of spoke bar. There is a strong electronic field in the center region of the spoke bar, while a high magnetic field around the loft areas of spoke bar and the connection region between the spoke base and cavity shell.

Dimensions of spoke thickness (T), spoke width (W) and spoke race-track height (A1) are optimized to minimize the ratio of $E_{peak}/E_{acc}$. With the change of T parameter, $E_{peak}/E_{acc}$ will present a lowest value around the ratio $T/L_{iris}$ of 0.3. Here, iris length is considered as a constant of 73mm. On the other hand, when the spoke width increases, the $E_{peak}/E_{acc}$ will go down to bottom and then go up. The R/Q value decreases with the increase of W dimension. The results of the simulations are shown in Figure 3.

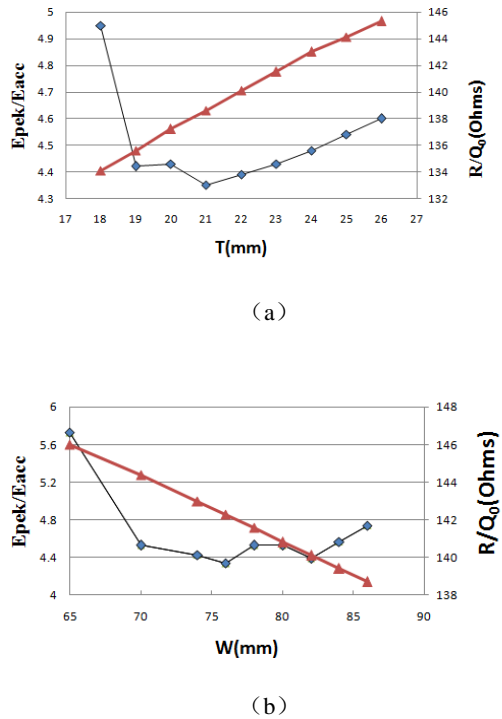

（a）

（b）

Figure 3: (a)Epeak/Eacc (blue)and R/Q0 （red）vs. T (b)Epeak/Eacc (blue)and R/Q0 （red）vs. W

For a spoke cavity, the peak surface magnetic field plays a critical role in the limitation of maximum accelerating grade. To achieve a lower peak surface magnetic field of Spoke012 cavity, different shapes and dimensions of spoke base have been studied. Three different shapes of spoke base and the RF results are shown in Figure 4 and Table 1 respectively. Finally, compared with round and race-track, elliptical spoke base has been chosen due to the more uniform magnetic field distribution and the higher shunt impedance.

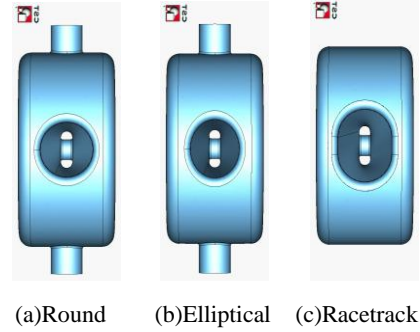

(a)Round  (b)Elliptical  (c)Racetrack

Figure 4: different shapes of spoke base

Table 1: RF parameter of different shapes of spoke base

| Spoke shape | round | elliptical | racetrack |
|---|---|---|---|
| G (Ω) | 51 | 53 | 70 |
| R/Q0 (Ω) | 124 | 135 | 93 |
| $E_{peak}/E_{acc}$ | 5.1 | 4.8 | 5.6 |
| $B_{peak}/E_{acc}$ (mT/(MV/m)) | 7.7 | 6.9 | 6.6 |

Here, $E_{acc}$ is defined as the total accelerating voltage divided by $\lambda *\beta$ .

Also the optimization of the ratio $D1/L_{cav}$ and spoke base diameters (D2 / D1) are investigated. As the D1 dimension goes up, $B_{peak}/E_{acc}$ and R/Q decrease in the same direction. In order to balance the higher R/Q with a lower $B_{peak}/E_{acc}$, the ratio $D1/L_{cav}$ of 0.5 is chosen. The influence of the $B_{peak}$ as the D2/D1 parameter changes must be studied. For an easy fabrication, ratio of 1.25 is preferred. The variation of $B_{peak}/E_{acc}$ and R/Q are shown in Figure 5.

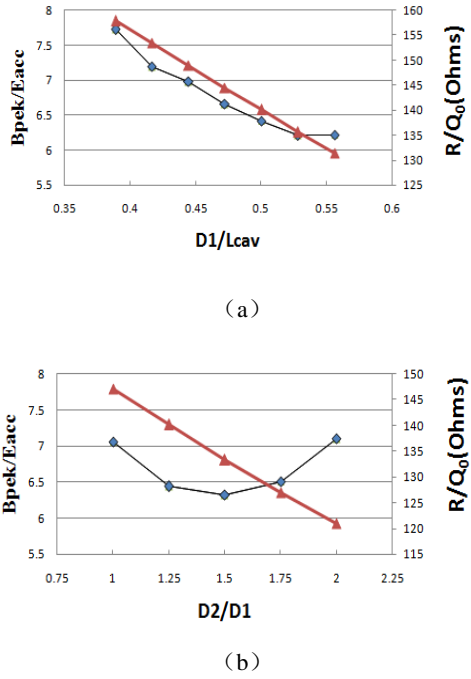

（a）

（b）

Figure 5: (a) $B_{peak}/E_{acc}$(blue) and R/Q0(red) results depend to the ratio $D1/L_{cav}$ (b) $B_{peak}/E_{acc}$(blue) and R/Q0(red) results depend to the ratio D2/D1

In addition, study on the influence of different end-wall shapes has been done because of the inherently poor mechanical stability of low Beta spoke cavity against helium pressure fluctuations. A camber end-wall has been selected instead of the traditional flat one for a better RF performance and stronger structure. Contrasted with a flat end-wall, the camber end-wall encloses the RF region with entire curves. One great advantage of the curved design is its ability to decentralize the helium pressure. It improves the mechanical property and is easier to design the stiffeners.

Comparisons of flat and camber end-walls are shown in Table 2.

### 3. Final geometry

After the optimization of all typical parameters, the main geometric parameters and RF results of spoke012 cavity are shown in table 3, meanwhile the field distribution are presented in Figure 6.

Table 2: Comparisons of flat and camber end-walls

| parameters | Flat end-wall | Camber end-wall |
|---|---|---|
| $E_p/E_{acc}$ | 4.8 | 4.4 |
| $B_p/E_{acc}$ | 7.0 | 6.3 |
| G | 53 | 61 |
| R/Q | 135 | 140 |
| df/dP (pipe free) | Peak stress =212 MPa -18 kHz/torr | Peak stress =157 MPa -10.9 kHz/torr |
| Tuning sensitivity | 1.3MHz/mm 776.9 kHz/kN | 1.0 MHz/mm 399.9 kHz/kN |
| The static Lorentz coefficient $Hz/(MV/m)^2$ | -25.8 | -15.1 |

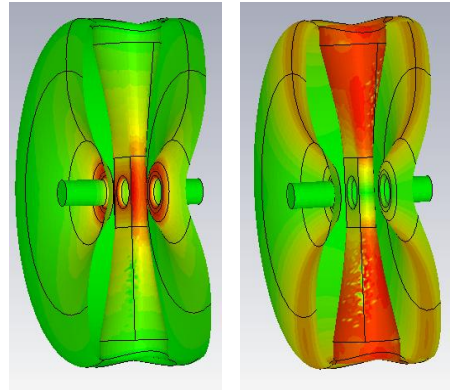

Figure 6：Surface electric field (left) and magnetic field (right), the field increases with the color

### 4. Mechanical design

In a continuous wave machine, the mechanical optimization focuses on the perturbations caused by the fluctuations of the liquid helium in the vessel. The exterior pressure results in the reduction deformation on the cavity volume. Volume reductions in the regions of high electric fields cause a negative frequency shift while reductions in the high magnetic field areas could have a positive shift. In the structure design of Spoke012 cavity, conventional stiff ribs are added in order to minimize the frequency shift and von stress due to the helium pressure vibration. Soildworks and Mechanical ANSYS software are used in these simulations.

Table 3: main geometric parameters and RF changing from green to red.

| | |
|---|---|
| Cavity length($L_{cav}$)/mm | 180mm |
| Cavity diameter($D_{cav}$)/mm | 468mm |
| Iris-to-iris length($L_{iris}$)/mm | 73mm |
| Major axis of spoke base (D2)/mm | 112.5mm |
| Minor axis of spoke base (D1)/mm | 90mm |
| Spoke thickness at aperture(T)/mm | 22mm |
| Spoke width at aperture(W)/mm | 82mm |
| Spoke race-track higher(A1)/mm | 94mm |
| Spoke lofting higher(A2)/mm | 398mm |
| Aperture diameter/mm | 30mm |
| Coupler port diameter/mm | 80mm |
| Operating frequency | 325MHz |
| Operating temperature in ADS | 4.2K |
| Accelerating gradient of Spoke012,$E_{acc}$ | 6MV/m |
| Q0 at nominal accelerating gradient | $5 \times 108$ |
| $E_{peak}/E_{acc}$ | 4.5 |
| $B_{peak}/E_{acc}$ | 6.4 mT/(MV/m) |
| G | 63Ω |
| R/Q | 142Ω |
| Geometrical Beta | 0.12 |
| Leff=β λ | 0.11m |

A view of stiffeners is shown in Figure 7.The structure of Spoke012 cavity is enhanced by three types of stiffeners. Two circular ribs with a thickness of 8mm in the end-wall outer region and six daisy ribs (6mm thickness)in the inner region aim to reduce the longitudinal deformation of the structure. A series of holes is presented on the joint of two circular stiffeners allowing the installation with the VT test fixture facility. To make sure the center shell is a full circle in the fabrication, four circumferential ribs on the cylindrical portion of the cavity are designed. Also this type of stiff ribs can lower the Lorentz forces. All stiffeners are made of reactor-grade niobium.

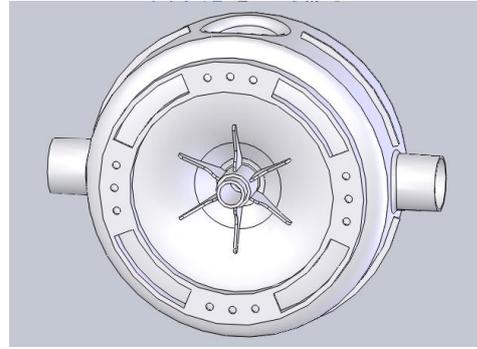

Figure 7 :stiffener design of Spoke012 cavity

A simulation result is shown in the Figure 8, where the deformation is due to exterior helium pressure of 1 atm. The design of the helium vessel was improved to reduce the sensitivity of the cavity to fluctuations to meet the requirement of df/dp. A bellow connecting with the beam flange on the tuner side can be seen in Figure 8. By optimizing the radius of circular ribs and bellows, the resultant force due to a pressure fluctuation at high electric field area near the beam pipe could be reduced. These simple approaches can decrease the sensitivity of helium pressure (df/dp) from -156Hz/torr to 17Hz/torr. In the calculation of external pressure loading, one beam pipe flange is welded to the helium vessel and is considered as "hard fixing". "Pipe free" means unconstrained at the bellow side, while "pipe fixed" means the tuner giving a longitudinal constraints about 170Kgf pre-tuning force. Helium vessel of Spoke012 is made of Titanium(TA2) .

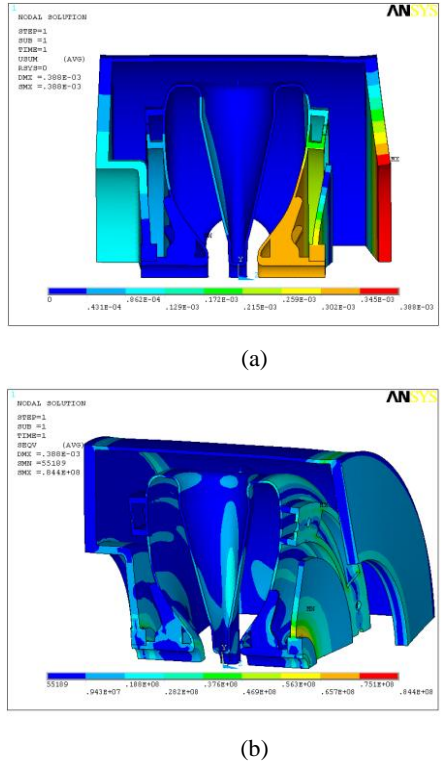

(a)

(b)

Figure 8: results of mechanical analysis with pre-tuning under 1atm (a)with ~170kgf pre-tuning by tuner, the maximum deformation on the cavity end wall where mainly a high magnetic field region (b)the von stress of cavity due to 1atm exterior pressure is under 30MPa.It is considered that Spoke012 will be safe under 2bar

Table 4 shows the mechanical parameters of the Spoke012 with its vessel. Volume shrinkage due to the temperature change is another typical reason of resonator frequency shift. The frequency increases about 460kHz when the cavity is cooled down from room temperature to 4.2K for Spoke012. The cavity is designed to be tuned only from the bellow side. With a tuning sensitivity of 1MHz/mm, Spoke012 cavity meets the requirement for the tuning range of 200kHz easily.

Table 4 mechanical parameters of the Spoke012

|  | With vessle |
|---|---|
| Pipe free | Von stress<40MPa $\Delta f / \Delta p = -156 Hz / torr$ |
| Pipe fixed | Von stress<50MPa $\Delta f / \Delta p = +17 Hz / torr$ |
| Tuner sensitivity | 60kHz/100kgf 1MHz/mm |
| The static Lorentz coefficient $Hz/(MV/m)^2$ | $-1.3 Hz/(MV/m)^2$ |
| Cooling down （300K-4.2K） | $\Delta f = +463 kHz$ |

## 4. Conclusion

Spoke012 is the first superconducting spoke resonator developed for China-ADS. The very low geometer Beta make it to be a key challenge. The RF design of the Spoke012 is finished, with the ratio of $E_p/E_{acc}$ is 4.5 and $B_p/E_{acc}$ is 6.4mT (MV/m). The structure of helium vessel is investigated to improve the mechanical performances. First two prototype cavities have been fabricated and cool tested. RF and mechanical performance were consistent with the simulation. Test results indicate that the design of the prototype cavity is successful and may meet the requirements of China-ADS.